\documentclass[graybox]{svmult}

\usepackage{amsmath}
\usepackage{amsfonts}
\usepackage{amssymb,bbm,units}

\usepackage{enumerate}

\usepackage{mathptmx}       
\usepackage{helvet}         
\usepackage{courier}        
\usepackage{type1cm}        
\usepackage{makeidx}         
\usepackage{graphicx}        
\usepackage{multicol}        
\usepackage[bottom]{footmisc}

\usepackage{
            ulem		
} 

\usepackage{pifont}
\usepackage{hyperref}

\newtheorem{assumption}[theorem]{Assumption}

\newcommand{\bE}{\mathbb{E}}
\newcommand{\bF}{\mathbb{F}}

\newcommand{\bN}{\mathbb{N}}
\newcommand{\bP}{\mathbb{P}}

\newcommand{\bR}{\mathbb{R}}

\newcommand{\cA}{\mathcal{A}}

\newcommand{\cF}{\mathcal{F}}

\newcommand{\cZ}{\mathcal{Z}}

\definecolor{darkgreen}{rgb}{0,0.35,0}

\newcommand{\geqs}{\geqslant}
\newcommand{\leqs}{\leqslant}

\makeindex             

\begin{document}

\title*{Forward utilities and Mean-field games under relative performance concerns\thanks{The authors express sincerest gratitude to T.~Zariphopoulou (University of Texas, US), M.~Anthropelos (University of Piraeus, GR), M.~Mrad (Universit\'e Paris 13, FR) for the helpful discussions and the two referees whose comments led to non-trivial improvements of the initial text.}}
\titlerunning{Forward utilities for many player games and Mean-field games }
\author{Gon\c calo dos Reis \thanks{G.~dos Reis acknowledges support from the \emph{Funda{\c c}$\tilde{\text{a}}$o para a Ci$\hat{e}$ncia e a Tecnologia} (Portuguese Foundation for Science and Technology) through the project  UIDB/00297/2020 (Centro de Matem\'atica e Aplica\c c$\tilde{\text{o}}$es CMA/FCT/UNL)}
and Vadim Platonov
}
\institute{Gon\c calo dos Reis \at The University of Edinburgh, School of Mathematics, Mayfield Road, The King's Buildings, Edinburgh, EH9 3FD, United Kingdom, and \\ Centro de Matem\'atica e Aplica\c c$\tilde{\text{o}}$es (CMA), FCT, UNL, Quinta da Torre,
2829-516 Caparica, Portugal, \email{G.dosReis@ed.ac.uk}
\and Vadim Platonov \at The University of Edinburgh, School of Mathematics, Mayfield Road, The King's Buildings, Edinburgh, EH9 3FD, United Kingdom, \email{v.d.platonov@sms.ed.ac.uk}
}
%
%
\maketitle

\abstract{We introduce the concept of mean field games for agents using Forward utilities of CARA type to study a family of portfolio management problems under relative performance concerns. 
Under asset specialization of the fund managers, we solve the forward-utility finite player game and the forward-utility mean-field game. We study best response and equilibrium strategies in the single common stock asset and the asset specialization with common noise. As an application, we draw on the core features of the forward utility paradigm and discuss a problem of time-consistent mean-field dynamic model selection in sequential time-horizons.   
}

\keywords{Forward utility, Mean-Field Games, social interactions, performance concerns.}
%
%
%

\section{Introduction}

This work brings together the concept of forward utilities to the mean-field game setting in the limelight of competitive optimal portfolio management of agents under relative performance criteria and the analysis of the associated finite-player game.

There exists a very rich literature on portfolio management for agents with utility preferences and under performance concerns to which this short introduction cannot possibly due justice. For a literature perspective of the financial setting including an in-depth discussion of agents with performance concerns and its impact in the utility maximization framework we refer to \cite{EspinosaTouzi2015,FreiDosReis2011,BielagkLionnetDosReis2017,deng2020relative} and references therein. Additionally, we point the reader to the beautiful introductions of \cite{LackerZariphopoulou2017,lackersoret2020many} where those concepts are brought to the framework of mean-field games.  Further, those works also make for an excellent review of mean-field games in the context of the Merton problem which is the framework underlying our work.

In short, mean-field games (MFG), stochastic or not, gained renewed interest due to their modelling power in crucially reducing the dimensionality  of the underlying problem under the assumption of statistically equivalent populations \cite{HuangMalhameCaines2006,cardaliaguet2015master,CarmonaDelarueLacker2017}. In other words, as long as the actions of a single agent do not affect the average interaction of the agents in their whole, then, in principle, the MFG framework stands to be more tractable than the $n$-agent games. See \cite{LackerZariphopoulou2017,lackersoret2020many}.

The novelty of our work is the conceptualization and analysis, simplified here, of the formulation of mean-field games within the so-called \textit{forward utilities} framework. Further, we juxtapose our construction to the related finite-player game. 

The classical and ubiquitous approach of \textit{utility preferences}, found throughout the literature \cite{EspinosaTouzi2015,FreiDosReis2011,BielagkLionnetDosReis2017,deng2020relative}, is that each agent, at an initial-time, specifies their risk-preferences to some future time $T$ and proceeds to optimize their investment to that initial-time. This \textit{backward} approach lacks flexibility to handle mid-time changes of risk-preferences by the agents, or, to allow an update of the underlying model: having in mind Covid-19, if the fund manager made investments in early 2019 to mature in the later part of 2020, how would one update the underlying model stock model to the change of parameters?

These problems feature an inherently \textit{forward-in-time} nature of investment. A view that is particularly clear for (competitive) fund managers updating their investment preferences frequently depending on market behavior. To cope with the limitation of the backward-in-time view induced by the classical utility optimization formulation, and, to better address this forward view, the mathematical tool of \textit{forward utilities} was developed. It was initially introduced for the analysis of the portfolio management problems in   \cite{MusielaZariphopoulou2007,MusielaZariphopoulou2008-TomKurzFestschrift,MusielaZariphopoulou2009} and subsequently expanded \cite{Zitkovic2009,Anthropelos2014,GechunZariphopoulou2016ergodic} and \cite{KarouiMrad2013,KarouiHillairetMrad2018,KarouiHillairetMrad2019}. The latter dealing with general forward utility It\^o random fields and with applications to longevity risk. Our approach builds from \cite{geng2017passive} where the first forward-utility definition under competition appeared (for finite-player games); we additionally refer the reader to the forthcoming \cite{AnthropelosGengZariphopoulou2020} (who also builds from \cite{geng2017passive}).  

In essence, the concept of forward utility reflects that the utility map must be adaptive and adjusted to the information flow. The forward \textit{dynamic} utility map is built to be consistent with respect to the given investment universe and the approach we discuss here is based on the martingale optimality principle (see Section \ref{sec:ClassicForwardUtility}).

To the MFG context, the closest to our work we have found is the concept of Forward-Forward MFG concept of \cite{GomesNurbekyanSedjro2016ForwardForward}.

\smallskip
\textbf{Organization of the paper.} 
In Section \ref{sec2} we introduce the financial market. In Sections \ref{sec:FRPCwithaverage} and \ref{sec:MFG-forward} we study the finite-agent and mean-field game respectively. We study forward utilities of time-monotone type. In Section \ref{sec:dynamicselection} we discuss the mean-field investment problem with dynamic model selection in large time-horizons.  We conclude in Section \ref{sec:Outlook} with a discussion of open questions and future research.

%
%
\section{Asset specialization, Forward utilities and CARA preferences}
\label{sec2}

\textbf{The market.} We consider a market environment with one riskless asset and $n$ risky securities which serve as proxies for two distinct asset classes. We assume their prices to be of log-normal type, each driven by two independent Brownian motions. More precisely the price $(S^i_t)_{t\geqslant 0}$ of the stock $i$ traded exclusively by the $i$-th agent solves 
\begin{align}
\label{stock}
\frac{dS_{t}^{i}}{S_{t}^{i}} 
& 
=\mu _{i}dt+\nu _{i}dW_{t}^{i}+\sigma _{i}dB_{t},
\end{align}
with constant parameters $\mu^i>0$, $\sigma_i\geqs 0$ and $\nu_i\geqs 0$ with $\sigma_i+\nu_i>0$. We refer the reader to \cite{LackerZariphopoulou2017,lackersoret2020many} for an in-depth motivation of the model. The one-dimensional standard Brownian motions $B,W^1,\cdots,W^n$ are independent. When $\sigma_i > 0$, the process $B$ induces a correlation between the stocks, and thus we call $B$ the \emph{common noise} and $W^{i}$ an \emph{idiosyncratic noise}. The independent Brownian motions $B,W^1,\cdots,W^n$ are defined on a probability space $(\Omega, \bF, \cF, \bP)$ endowed with the natural filtration $\bF=(\cF_t)_{t\geqs 0}$ generated by them and satisfies the usual conditions. 

We recall the case of \textit{single common stock}, where for any $i  = 1,\dots,n ,~ (\mu_i,\sigma_i) = (\mu,\sigma),~ \nu_i = 0,$ for some $\mu, \sigma >0$ and independent of $i$. The single common stock case has been explored in great generality in \cite{EspinosaTouzi2015,FreiDosReis2011,BielagkLionnetDosReis2017} incorporating portfolio constraints, general stock price dynamics and risk-sharing mechanisms.

We aim to contribute to the literature on mean field
games and forward utilities by providing an explicitly solvable example. As argued by \cite{lackersoret2020many}, outside linear-quadratic structures such is very rare, and it is one of these rarities we bring here. We work with the very tractable model \eqref{stock} and include common noise, heterogeneous of agents, a mean field interaction through the controls in addition to the state processes and forward utilities.

\textbf{Agents' wealth.} Each agent $i=1,\ldots,n$ trades using a self-financing strategy, $(\pi _{t}^{i})_{t \geqs 0}$, which represent the (discounted by the bond) amount invested in the $i$-th stock. The $i^{th}$ agent's wealth $(X^i_t)_{t\geqs 0}$ then solves 
\begin{equation}
\label{def:agentsWealthX-i}
dX_{t}^{i} 
=
 \pi_t^i\Big(\mu _{i}dt+\nu _{i}dW_{t}^{i}+\sigma _{i}dB_{t}\Big), \quad \textrm{with}\quad X_0^i=x_0^i \in \bR.
\end{equation}
We recall that the strategy is self-financing, when the agent wealth evolve from the starting capital only by agent's investment decisions in the market without any external sources of income and this evolution is described by respective SDE \eqref{def:agentsWealthX-i}.

A portfolio strategy is said admissible if it belongs to the set $\cA^i$, which consists of
\begin{align*}
\cA^i 
= 
\Big\{
\pi^i&: 
\textrm{$\bF$-progressively measurable $\bR$-valued processes } (\pi^i_t)_{t \geqs 0},
\\
& \qquad 
\textrm{and self-financing such that }
\bE[\int_0^t |\pi_s|^2 ds ]< \infty,~ \text{for all } t \geqs 0
\Big\}.
\end{align*}
\textbf{The Agents' social interaction.}
Each manager measures the performance of her strategy taking into account the policy of the other. Each agent engages in a form of social interaction that affects the agent's perception of wealth, all in an additive fashion modelled through the arithmetic average wealth of all agents (this model is largely inspired in \cite{EspinosaTouzi2015,FreiDosReis2011,BielagkLionnetDosReis2017,LackerZariphopoulou2017}). The way the agent assesses and optimizes his relative performance is explored through Definition \ref{def:ForwardUtilityBestResponse} in the latter Section \ref{sec:FRPCwithaverage}. So far we introduce the \emph{relative performance metric} of manager $i\in\{1,\dots,n\}$, denoted $\widetilde X^i$ is defined to be
\begin{align}
\label{eq:xminusaverage}
 \widetilde X^i = X^i - \theta_i\overline X,
\quad\textrm{where}\quad
\overline{X}:=\frac{1}{n}\sum_{k=1}^{n}X^{k}
\quad\textrm{and}\quad \theta_i\in[0,1],  
\end{align}
where deterministic $\theta_i$ stands for the competition weight for agent $i$. 

We easily obtain a dynamics for $\overline X$ and $\widetilde X^i$, namely
\begin{align}
\nonumber
d\overline X_{t} 
&=\Big(\frac1n\sum_{k=1}^n \pi_t^k \mu_{k}\Big)dt
+\Big(\frac1n\sum_{k=1}^n \pi_t^k \nu_{k}dW_{t}^{k}\Big)
+\Big(\frac1n\sum_{k=1}^n \pi_t^k \sigma_{k} \Big)dB_{t}
\\
\nonumber
&
= \overline{(\pi \mu)}_t dt
+\Big(\frac1n\sum_{k=1}^n \pi_t^k \nu_{k}dW_{t}^{k}\Big)
+ \overline{(\pi \sigma)}_t dB_t,\quad \overline X_0 = \overline x_0 = \frac1n\sum_{k=1}^n x_0^k
\\
\nonumber
d\widetilde X^i_t
&
=\big(\pi_t^i \mu _{i}- \theta_i\overline{(\pi \mu)}_t \big)dt +\Big(\pi_t^i\nu _{i}dW_{t}^{i}-\theta_i\big(\frac1n\sum_{k=1}^n \pi_t^k\nu _{k}dW_{t}^{k}\big)\Big)
\\
\label{eq:WidetildetXi}
&\qquad\qquad
+\big(\pi_t^i\sigma _{i}- \theta_i\overline{(\pi \sigma)}_t \big)dB_t,\quad \widetilde X^i_0 =  x^i_0-\theta_i\overline x_0, 
\end{align}
where $\overline x_0$, $\overline{\pi \mu}$ and $\overline{\pi \sigma}$ are identified as averages (as seen from the 1st equation to the 2nd). Similarly to \cite[{Remark 2.5}]{LackerZariphopoulou2017}, it is natural to replace the average wealth $\overline X$ in \eqref{eq:xminusaverage} by the average over all other agents. With that in mind we define for convenience $\overline{X}^{(-i)} = \frac{1}{n-1}\sum_{k \neq i}X^k$ and $Y^{(-i)}=\frac{n}{n-1}\overline{X}^{(-i)}$. This leads us to recast \eqref{eq:xminusaverage} as
\begin{align}
\label{eq:xminusaverage-lessi}
 \widehat X^i = X^i - \theta_i\overline{X}^{(-i)},
\qquad\textrm{where}\qquad
\overline{X}^{(-i)} = \frac{1}{n-1}\sum_{k \neq i}X^k.  
\end{align}
We easily obtain a dynamics for $\widehat X$ and $\overline{X}^{(-i)}$, namely
\begin{align}
\nonumber
d\overline{X}^{(-i)}_{t} 
&
= 
\overline{(\pi \mu)}^{(-i)}_t dt
+\Big(\frac1{n-1}\sum_{k\neq i} \pi_t^k\nu_{k}dW_{t}^{k}\Big)
+ \overline{(\pi \sigma)}^{(-i)}_t dB_t,\quad \overline{X}^{(-i)}_0 = \overline x_0^{(-i)}
\\
\nonumber
d\widehat X^i_t
&
= \big(\pi_t^i \mu _{i}- \theta_i\overline{(\pi \mu)}^{(-i)}_t \big)dt
+\Big(\pi_t^i\nu _{i}dW_{t}^{i}-\theta_i\big(\frac1{n-1}\sum_{k\neq i}^n \pi_t^k\nu_{k}dW_{t}^{k}\big)\Big) \\
\label{eq:WidehatXi}
&\qquad
+\big(\pi_t^i\sigma _{i}- \theta_i\overline{(\pi \sigma)}^{(-i)}_t \big)dB_t,\quad \widehat X^i_t =  x^i_0-\theta_i\overline x_0^{(-i)} .
\end{align}
We also define the quantities 
\begin{align*}
\widehat{\pi \sigma}^{(-i)}:=\frac1n \sum_{k\neq i}\pi^k  \sigma_k,\ \ \
\overline{(\pi \mu)}^{(-i)}:=\frac1n \sum_{k\neq i}\pi^k  \mu_k\ \ \textrm{and}\ \  
\overline{(\pi \nu)^2}^{(-i)}:=\frac1n \sum_{k\neq i}(\pi^k  \nu_k)^2,
\end{align*}
where we have the following relations between $\widehat{\pi \sigma}^{(-i)}$, $\overline{\pi \sigma}^{(-i)}$ and $\overline{\pi \sigma}$:
\begin{align}
 \label{eq:identitiesBetweenAverages}
	\overline{\pi \sigma}^{(-i)} 
	&= \frac n{n-1} \overline {\pi \sigma} - \frac1{n-1} \pi^i \sigma_i
	,\quad
	\overline{\pi \sigma}^{(-i)}  = \frac{n}{n-1} \widehat{\pi \sigma}^{(-i)},
\end{align}
and $\widehat{\pi \sigma}^{(-i)} = \overline {\pi \sigma} - \frac 1n \pi^i \sigma_i$. We do not write it explicitly but we extend the same notation and relations to $\widehat{\pi \mu}^{(-i)}$, $\overline{\pi \mu}^{(-i)}$ and $\overline{\pi \mu}$.

\subsection{Forward dynamic utilities (classic)}
\label{sec:ClassicForwardUtility}

We recall, for reference, the classic \textit{forward utility} formulation. We define a \emph{forward dynamic utilities} in the context of the probability space $(\Omega, \bF, \cF, \bP)$. We denote by $u_0:\bR\to \bR$ the initial data. The forward utility is constructed based on the martingale optimality principle.

\begin{definition}[Forward dynamic utilities]
Let $U:\Omega \times\bR\times [0,\infty)\to \bR$ be an $\bF$-progressively measurable random field. $U$ is a \emph{forward dynamic utility} if 
\begin{itemize}
	\item For all $t\geqs 0$ the map $x\mapsto U(x,t)$ is $\bP$-a.s.~increasing and concave;
	\item It satisfies $U(x,0)=u_0(x)$;
	\item For all $T\geqs t$ and each self-financing strategy, represented by $\pi$, the associated discounted wealth process $X^\pi$ satisfies a supermartingale property
	\begin{align*}
		\bE[ U(X^\pi_T,T)|\cF_t] & \leqs U(X^\pi_t,t) \quad \bP\textrm{-a.s.};
	\end{align*}
	\item For all $T\geqs t$ there exists a self financing strategy, represented by $\pi^*$, for which the associated discounted wealth $X^*$ satisfies a martingale property
		\begin{align*}
		\bE[ U(X^*_T,T)|\cF_t] & = U(X^*_t,t) \quad \bP\textrm{-a.s.}
	\end{align*}
\end{itemize}
\end{definition}
The above definition assumes the optimizer is attained. This is a somewhat strong assumption which is discussed in  \cite{Zitkovic2009,Anthropelos2014}. There it is argued that such constraint is not necessary for the forward utility construction in certain contexts. 

Following e.g.~\cite[{Section 5}]{MusielaZariphopoulou2009}, we say a utility map $U$ is of \textit{Constant Absolute Risk Aversion} (CARA) type if the \emph{local risk tolerance function} $r$, given by the quotient $r(\cdot)=-U_x(\cdot)/U_{xx}(\cdot)$, is constant uniformly. This is the case for the classical exponential utility function, see Example \ref{Example:ExpUtility} below.

%
%
%
%
%

%
%
%
%
%
\section{Forward relative performance criteria}
\label{sec:FRPCwithaverage}

\subsection{Forward relative performance criteria}
\label{sec:FRPCwithaverage-n-1}

Each manager measures the output of her relative performance metric using a forward relative one as modelled by an $\cF_t$-progressively measurable random field $U^i:\bR\times [0,\infty) \to \bR$ for $i\in\{1,\dots,n\}$. The below criteria follows those proposed in \cite{geng2017passive}.

The main idea here being a formulation inspired in the first step in the usual strategy of solving a Nash game, namely the best response of an agent to the actions of all other agents. Take manager $i$ and assume all other agents $j\neq i$ have acted with an investment policy $\pi^j$ then for any strategy $\pi^i\in \cA^i$, the process $U^i(\widehat X^i_t,t)$ is a (local) supermartingale, and there exists $\pi^{i,*}\in \cA^i$ such that $U^i(\widehat X^{i,*}_t,t)$ is a (local) martingale where $\widehat X^{i}$ and $\widehat X^{i,*}$ solves \eqref{eq:xminusaverage-lessi} with strategies $\pi^i$ and $\pi^{i,*}$ respectively. 

This version of a relative criterion is (implicitly and) exogenously parametrized by the policies of all other managers $j\neq i$ over which there is no assumption on their optimality. In Nash-game language, we solve the so-called best response.
\begin{definition}[Forward relative performance for the manager]
\label{def:ForwardUtilityBestResponse}
Each manager $i\in\{1,\cdots,n\}$ satisfies the following. Let $\pi^j\in\cA^j$, for any $j\neq i$ be arbitrary but fixed admissible policies, in other words, the other managers have fixed their admissible strategies. 

An $\bF$-progressively measurable random field $U^i(x,t)$  is a \emph{forward relative performance} for manager $i$ if, for all $t\geqs 0$, the following conditions hold:
\begin{enumerate}[i)]
	\item The mapping $x \mapsto U^i(x,t)$, 
	is $\bP$-a.s.~strictly increasing and strictly concave;
	\item  For any $\pi^i \in \cA^i$, $U^i( \widehat X^i_t,t)$ is a (local) supermartingale and $\widehat X^i$ is the relative performance metric given in \eqref{eq:xminusaverage-lessi};
	\item There exists $\pi^{i,*}\in \cA^i$ such that $U^i(\widehat X^{i,*}_t,t)$ is a (local) martingale where $\widehat X^{i,*}$ solves \eqref{eq:xminusaverage-lessi} with strategies $\pi^{i,*}$ being used.  
\end{enumerate}
\end{definition}
In the above definition, we do not make explicit references to the initial conditions $U^i(x,0)$ but we assume that admissible initial data
exists such that the above definition is viable. Contrary to the classical expected
utility case, the forward utility process is an investor-specific input. Once it 
is chosen, the supermartingale and martingale properties impose conditions
on the drift of the process. Under enough regularity, these conditions lead to
the forward performance SPDE (see \cite{MusielaZariphopoulou2010}).

Since we are working in a log-normal market, it suffices to study smooth relative performance criteria of zero volatility (of the forward utility map). Such processes are extensively analysed in \cite{MusielaZariphopoulou2010-space-time-monotone} in the absence of relative performance concerns. There, a concise characterization of the forward criteria is given along necessary and sufficient  conditions for their existence and uniqueness. In that setting, the zero-volatility forward processes are always time-decreasing processes. We point to the reader that this does not have to be case if relative performance concerns are present (see also \cite{geng2017passive}).
Before proving the main result of the subsection, we make a standing assumption regarding the regularity of the forward utility maps

\begin{assumption}
\label{ass:regularity-u-n-player}
Assume that the derivatives $U^i_t(x,t)$, $U^i_{x}(x,t)$ and $U^i_{xx}(x,t)$ exists for $t\geqs 0,~x\in\bR$, $\bP$-a.s.
\end{assumption}
From Assumption \ref{ass:regularity-u-n-player}, the It\^o decomposition of the forward utility map is
\begin{align}
\label{eq:SDEforForwardUtilityField}
d U^i(x,t)=U^i_t(x,t)dt,\quad \textrm{for }i\in\{1,\cdots,n\}.
\end{align}

We next derive a PDE with random coefficients and an optimal investment strategy for a smooth relative performance criteria of zero volatility of some agent $i$ assuming that all other agents $j\neq i$ have made their investment decisions.

\begin{proposition}[Best responses]
\label{prop:BestResponses-n-playerGame-average-n-1}
Fix $i\in\{1,\dots,n\}$ and the agent's initial preference $u^i_0$. Assume that each manager $j\neq i$ follows $\pi^j\in\cA^j$. Consider the PDE with stochastic coefficients for $(x,t)\in \bR\times [0,\infty)$
\begin{align}
\label{eq:SPDE-step01}
\nonumber
U^i_t& =\Big(\theta_i\overline{(\pi \mu)}^{(-i)}_t - \frac{\mu_i\theta_i \sigma_i \overline{(\pi \sigma)}^{(-i)}_t}{\nu_i^2+\sigma_i^2} \Big)U^i_x 
       + \frac{\mu^2_i}{2(\nu_i^2+\sigma_i^2)} \frac{(U^i_x)^2}{U^i_{xx}}
\\
& \qquad\qquad 
  + \frac12 U^i_{xx} \Big[  \Big(\theta_i\overline{(\pi \sigma)}^{(-i)}_t\Big)^2 \Big( \frac{ \sigma_i^2 }{\nu_i^2+\sigma_i^2} -1 \Big)
	                       -\frac{\theta_i^2}{n-1}\overline{(\pi \nu)^2}^{(-i)}\Big],
\end{align}
and assume that for an admissible initial condition $U(\cdot,0)=u^i_0(\cdot)$, the PDE has a smooth solution $U^i$ satisfying Assumption \ref{ass:regularity-u-n-player}, such that $x\mapsto U^i(x,t)$ is strictly increasing ($U_x>0$) and strictly concave ($U_{xx}<0$) for each $t>0$ $\bP$-a.s.

Define the strategy $\pi^{i,*}$
\begin{align*}
\pi^{i,*}_t 
& 
=
\frac1{\nu_i^2+\sigma_i^2} \Big( {\theta_i \sigma_i \overline{(\pi \sigma)}^{(-i)}_t} 
- \mu_i\frac{U^i_x(\widehat X^{i,*}_t,t)}{U^i_{xx}(\widehat X^{i,*}_t,t)}\Big),\quad t>0,
\end{align*}
where $\widehat X^{i,*}$ solves \eqref{eq:WidehatXi} with $\pi^{i,*}$ being used.

If $\pi^{i,*}\in\cA^i$ and $\widehat X^{i,*}$ are well-defined, then $U^i(x,t)$ is a forward utility performance process. Moreover, the policy $\pi^{i,*}$ is optimal (in the sense of Definition \ref{def:ForwardUtilityBestResponse}).
\end{proposition}

\begin{remark}
Note that the randomness in PDE \eqref{eq:SPDE-step01} is coming from $\pi^\cdot$ only. 
\end{remark}

Using the language of \cite[{Section 5}]{MusielaZariphopoulou2009}, define the \emph{local risk tolerance function} $r^i:\Omega\times\bR\times [0,\infty)\to \bR$ such that $r^i(x,t):=-U^i_x(x,t)/U^i_{xx}(x,t)$. Then, by direct inspection of the expression for $\pi^{i,*}$ one sees that if the \textit{local risk tolerance function} $r^i(x,t)=r^i={Const}$, for all $t>0$ (e.g. the utility is of \textit{Constant Absolute Risk Aversion (CARA)} type -- see Section \ref{sec:ClassicForwardUtility}) then the optimal strategy will be constant throughout time if additionally all other agents also choose a constant strategy. 

\begin{corollary}[Constant strategies under CARA]
Assume that all agents $j\neq i$ invest according to constant strategies $\pi^j\in\bR$ and that the \emph{local risk tolerance function} $r^i$ is constant. Then $\pi^{i,*}$ is constant.
\end{corollary}

We now prove the previous ``best responses'' proposition above.
\begin{proof}[of Proposition \ref{prop:BestResponses-n-playerGame-average-n-1}]
From \eqref{eq:xminusaverage-lessi} we have the dynamics of $d\widehat X^i$ (and hence that of $d(X^i - \theta_i\overline{X}^{(-i)})$). We now apply the It\^o formula to $U^i(\widehat X^i_t ,t)=U^i(X^i_t - \theta_i\overline{X}^{(-i)}_t ,t)$,
\begin{align}
\nonumber
 d U^i(\widehat X^i_t ,t) 
&= U^i_t(\widehat X^i_t ,t)dt + U^i_x(\widehat X^i_t ,t) d \widehat X^i_t + \frac12 U^i_{xx}(\widehat X^i_t ,t) d \langle \widehat X^i_t \rangle 
\\ 
\nonumber
&
=U^i_t(\widehat X^i_t ,t)dt  
+ U^i_x(\widehat X^i_t ,t) \big(\pi_t^i \mu _{i}- \theta_i\overline{(\pi \mu)}^{(-i)}_t \big)dt
\\ 
\label{eq:generalSDEafterItoWentzell}
&
\quad
+ U^i_x(\widehat X^i_t ,t)\Big(\pi_t^i\nu _{i}dW_{t}^{i}-\theta_i\big(\frac1{n-1}\sum_{k\neq i}^n \pi_t^k\nu_{k}dW_{t}^{k}\big)\Big)\\
\nonumber
&
\quad 
+ U^i_x(\widehat X^i_t ,t)\big(\pi_t^i\sigma _{i}- \theta_i\overline{(\pi \sigma)}^{(-i)}_t \big)dB_t
\\
\nonumber
&
\quad 
+\frac12 U^i_{xx}(\widehat X^i_t ,t) 
\Big[ (\pi^i_t \nu_i)^2 
	    +\frac{\theta_i^2}{n-1}\overline{(\pi \nu)^2}^{(-i)}
      + \big(\pi_t^i\sigma _{i}- \theta_i\overline{(\pi \sigma)}^{(-i)}_t \big)^2
\Big]dt,
\end{align}
with $U^i(\widehat X^i_0 ,0)=U^i(x^i_0-\theta_i\overline x_0^{(-i)},0)$ and we used that the $B,W^j$ are all i.i.d.

By Definition \ref{def:ForwardUtilityBestResponse}, the process $U^i(\widehat X^i_t,t)$ becomes a Martingale at the optimum $\pi$. Direct computations using first order conditions ($\partial_{\pi^i} \textrm{``drift''}=0$) yield 
\begin{align}
\label{eq:optimstrategy}
\nonumber
 0 &+ U^i_x \big( \mu_i -0\big) + \frac12 U_{xx}^i\Big[2 \pi^i \nu_i^2 + 0 + 2\big(\pi_t^i\sigma _{i}- \theta_i\overline{(\pi \sigma)}^{(-i)}_t \big)\sigma_i \Big] =0
\\
&\Leftrightarrow \qquad  U_{xx}^i \pi^i (\nu_i^2+\sigma_{i}^2) =-U^i_x \mu_i + U_{xx}^i \theta_i \sigma_i \overline{(\pi \sigma)}^{(-i)}_t 
\\
\nonumber
&~\Rightarrow \qquad \pi^i_t 
=
\frac1{\nu_i^2+\sigma_i^2} \Big({\theta_i \sigma_i \overline{(\pi \sigma)}^{(-i)}_t} - \mu_i\frac{U^i_x(\widehat X^i_t ,t)}{U^i_{xx}(\widehat X^i_t ,t)}\Big). &
\end{align}
Injecting the expression of $\pi^i_t $ in the drift term of \eqref{eq:generalSDEafterItoWentzell} and simplifying we arrive at the consistency condition \eqref{eq:SPDE-step01}, we do not carry out this step explicitly, nonetheless, using that $U^i$ solves \eqref{eq:SPDE-step01} equation \eqref{eq:generalSDEafterItoWentzell} simplifies to (exact calculations are carried out in the Section \ref{sec:appendix}),
\begin{align}
\nonumber
& d U^i(\widehat X^i_t ,t)
\\ \nonumber
&
= U^i_x(\widehat X^i_t ,t)\Big(\pi_t^i\nu _{i}dW_{t}^{i}-\theta_i\big(\frac1{n-1}\sum_{k\neq i}^n \pi_t^k\nu_{k}dW_{t}^{k}\big)\Big)
\\
\nonumber
\label{eq:SDEafterItoWentzell and choice of Ut-average-n-1}
&
 \quad
+ U^i_x(\widehat X^i_t ,t)\Big(\pi_t^i\sigma _{i}- \theta_i\overline{(\pi \sigma)}^{(-i)}_t \Big)dB_t
\\
&
 \quad 
+\frac12 U^i_{xx}(\widehat X^i_t ,t)\frac1{\nu_i^2+\sigma_i^2}  
\Big| {\pi^i}(\nu_i^2+\sigma_i^2) -\Big( \theta_i \sigma_i \overline{(\pi \sigma)}^{(-i)}_t - \mu_i\frac{U^i_x(\widehat X^i_t ,t)}{U^i_{xx}(\widehat X^i_t ,t)} \Big)\Big|^2dt.
\end{align}
The concavity assumption of $U^i(x,t)$ implies that the drift term above is non-positive and vanishes when \eqref{eq:optimstrategy} holds. We can conclude that, if $\pi^{i,*}_t=\pi^i_t\in \cA^i$ and the associated process $\widehat X^{i,*}$ is well-defined (solution to \eqref{eq:WidehatXi} with $\pi^{i,*}$), the process $U^i(\widehat X^{i,*}_t,t)$ is a local-martingale, otherwise it is a local supermartingale.
\end{proof}

\subsubsection{Examples: CARA case}

\begin{example}[The classic CARA case - exponential case]
\label{Example:ExpUtility}
The exponential criterion takes as initial condition the map $U(x,0)$ ($x\in\bR$) defined as 
\begin{align}
\label{eq:ExponentialUtility-timeZero} 
U^i(x,0) & = -e^{-x/\delta},\quad \textrm{with} \quad \delta>0.
\end{align}
In this case, the \emph{local risk tolerance function} $r=-U^i_x/U^i_{xx}=\delta$.
\end{example}

In our case accounting for social interaction between agents in the form of performance concerns, the $i$-th agent's utility is a function $U^i:\Omega\times \bR\times \bR \times [0,\infty)\to \bR$ of both her individual wealth $x$ and the average wealth wealth of all agents, $m$. The initial/starting utility map is of the form 
\begin{align*}
U^i(x,m,0) & = -\exp\Big\{ -\frac1{\delta_i} (x -\theta_i m)\Big\},
\end{align*}
where we refer to the constants $\delta_i>0$ and $\theta_i\in[0,1]$ as \emph{personal risk tolerance} and \emph{competition weight} parameters, respectively.

\begin{example}[The time-monotone forward utility with starting exponential]
\label{ex:ForwardExponentialUtility}
For $i\in\{1,\cdots,n\}$, let the dynamics of $U^i$ be given by \eqref{eq:SDEforForwardUtilityField} and assume $U^i(x,0) = -e^{-{x}/{\delta_i}}$ with $\delta_i>0$. Then the solution to the PDE \eqref{eq:SPDE-step01} is given by 
\begin{align}
\label{eq:Solution Postulation} 
U^i(x,t) & = -e^{-\frac x {\delta_i}+f_i(t)},\qquad \textrm{with} \quad {\delta_i}>0,
\end{align}
where $(f_i(t))_{t\geqs 0}$ is the random map given below independent of $x$ satisfying $f_i(0)=0$, sufficiently integrable and $t\mapsto f_i(t)$ is differentiable. Note that in this case, the \emph{local risk tolerance function} satisfies $r^i=-U^i_x/U^i_{xx}=\delta_i$.

Injecting $U^i(x,t)$ above in \eqref{eq:SPDE-step01} yields an ODE for $f_i$ (we omit the time variable), 
\begin{align*}
f_i'& =-\frac{ \theta_i}{\delta_i} \Big(\overline{(\pi \mu)}^{(-i)} - \frac{\mu_i \sigma_i \overline{(\pi \sigma)}^{(-i)}}{\nu_i^2+\sigma_i^2} \Big) 
       + \frac{\mu^2_i}{2(\nu_i^2+\sigma_i^2)}
\\
& \qquad\qquad 
  + \frac{\theta_i^2}{2\delta_i^2}  \Big[  \Big(\overline{(\pi \sigma)}^{(-i)}\Big)^2 \Big( \frac{ \sigma_i^2 }{\nu_i^2+\sigma_i^2} -1 \Big)
		                       -\frac{1}{n-1}\overline{(\pi \nu)^2}^{(-i)}\Big]
\\
&= -\frac{ \theta_i}{\delta_i} \overline{(\pi \mu)}^{(-i)}
+ \frac{1}{2(\nu_i^2+\sigma_i^2)}\Big( \mu_i + \frac{\theta_i}{\delta_i} \sigma_i \overline{(\pi \sigma)}^{(-i)} \Big)^2\\
&
\qquad \qquad - \frac{\theta^2_i}{2\delta_i^2}\Big[\Big(\overline{(\pi \sigma)}^{(-i)}_t\Big)^2 
+\frac{1}{n-1}\overline{(\pi \nu)^2}^{(-i)}\Big] =: \lambda_i.
\end{align*}

Hence, $f_i(t)=\int_0^t \lambda_i(s)ds$. In particular, if all coefficients and strategies are constant, then (with a slight abuse of notation) $f_i(t)=t \lambda_i $ for a constant $\lambda_i$ given by the RHS of the above ODE.
\end{example}

\begin{example}[No performance concerns: $\theta^i=0$]
We continue to work under the time-monotone forward utility case of the previous example. 
Without performance concerns, i.e.~$\theta_i=0$, then $\lambda_i$ is just the Sharpe ratio $\lambda_i=\frac{\mu^2_i}{2(\nu_i^2+\sigma_i^2)}$ and we recover well-known results. We have from Proposition \ref{prop:BestResponses-n-playerGame-average-n-1} that
\begin{align*}
\pi^{i,*}_\cdot = \frac{\mu_i \delta_i}{\nu_i^2+\sigma_i^2} 
\quad\textrm{and}\quad
U^i(x,t) = -\exp\Big\{ -\frac x {\delta_i} +t \lambda_i^{(\theta_i=0)}\Big\},
\end{align*}
with the constant $\lambda_i^{(\theta_i=0)}$ just being the Sharpe ratio, $\lambda_i^{(\theta_i=0)}=\frac{\mu^2_i}{2(\nu_i^2+\sigma_i^2)}$.
\end{example}

\subsection{The Forward Nash equilibrium}

In view of the \emph{best responses} discussed in Proposition \ref{prop:BestResponses-n-playerGame-average-n-1} we now investigate the \emph{simultaneous best responses} as to establish the existence of a Nash equilibrium. 
\begin{definition}[Forward Nash equilibrium]
\label{def:ForwardNashEquilibrium}
A forward Nash equilibrium consists of $n$-pairs of $\bF$-adapted maps $(U^i,\pi^{i,*})$ such that for any $t\geqs 0$ the following conditions hold.
\begin{itemize}
	\item For any $i\in \{1,\cdots, n\}, ~\pi^{i,*}\in\cA^i$;
	\item For each player $i\in\{1,\cdots,n\}$ the following holds: given the strategies $\pi^{j,*}\in\cA^j$ (any $j\neq i$) the processes $U^i(\widehat X^{i}_t(\pi^{*,-i}),t)$ is a (local) supermartingale where $\widehat X^{i}(\pi^{*,-i})$ solves \eqref{eq:WidehatXi} with all managers $j\neq i$ acting according to $\pi^{j,*}$;
	\item For each player $i\in\{1,\cdots,n\}$ the following holds: the process  $U^i(\widehat X^{i,*}_t(\pi^{*,-i}),t)$  is a (local) martingale where $\widehat X^{i}(\pi^{*,-i})$ solves \eqref{eq:WidehatXi} with \emph{all} managers $j$ acting according to $\pi^{j,*}$. 
\end{itemize}
If all the optimal strategies are constant we say we have a \emph{constant forward Nash equilibrium}.
\end{definition}
Under appropriate integrability conditions plus the martingale/supermartingale characterizations, we have for some agent $i$ for any $\pi^i\in\cA^i$ 
\begin{align*}
	\bE[ U^i(\widehat X^{i,*}_t(\pi^{*,-i}),t) ] 
	&= \bE[ U^i(\widehat X^{i,*}_0(\pi^{*,-i}),0) ]
	= \bE[ U^i(x^i_0-\theta_i\overline x_0^{(-i)},0) ]	
	\\
	&
	= U^i(x^i_0-\theta_i\overline x_0^{(-i)},0) 	
	\geqs \bE[ U^i(\widehat X^{i}_t(\pi^{*,-i}),t) ].
\end{align*}
As expected, no manager can increase the expected utility of her relative performance metric by unilateral decision.

The solvability of the general forward Nash equilibrium seems very difficult for a general forward criteria as one needs to solve the following system for the $\pi^{i,*}$ (see Proposition \ref{prop:BestResponses-n-playerGame-average-n-1}, in particular \eqref{eq:optimstrategy}) and the corresponding PDEs for the $U^i,~ i\in\{1,\cdots,n\}$:
\begin{align}
\label{eq:GeneralStrategySystem} 
\pi^{i,*}_t (\nu_i^2+\sigma_i^2)
& 
=
\theta_i \sigma_i  \Big(\frac1{n-1}\sum_{k=1,k\neq i}^n \pi^{k,*}_t \sigma_{k} \Big)
-  \mu_i\frac{U^i_x\big(\widehat X^{i,*}_t(\pi^{*,-i}),t\big)}{U^i_{xx}\big(\widehat X^{i,*}_t(\pi^{*,-i}),t\big)}.
\end{align}

\subsubsection{Equilibrium with time-monotone forward utilities and exponential initial condition}

In order to obtain explicit results we focus on the time-monotone case presented in Example \ref{ex:ForwardExponentialUtility} for which $U^i_x/U^i_{xx}=-\delta_i$. More notably,  at the level at which we have formulated our problem we can easily recover the results of \cite[{Theorem 2.3}]{LackerZariphopoulou2017} for which one has $U^i_x/U^i_{xx}=-\delta_i$, for any $t$ (note their Remark 2.5). 

\begin{theorem}
\label{theo:nPlayerForwardNashGame}
Assume the conditions of Proposition \ref{prop:BestResponses-n-playerGame-average-n-1} hold for all agents $i\in\{1,\cdots,n\}$. Assume furthermore that agents have time-monotone forward utility $U^i$ with initial condition \eqref{eq:ExponentialUtility-timeZero}.

Define the quantities  $\varphi^\sigma_{n}$ and $\psi^\sigma_{n}$ by
\begin{equation}
\varphi^\sigma_{n} :=
\frac1n\sum_{i=1}^n  \delta_i \frac{\mu_i \sigma_i }{{\nu_i^2 + \sigma^2_i \big(1+\frac{\theta_i}{n-1}\big) }}
\ \ \text{ and } \ \ 
\psi^\sigma_n := 
\frac1{n-1}\sum_{i=1}^n  \theta_i  \frac{\sigma^2_i}{{\nu_i^2+\sigma_i^2 \big(1+\frac{\theta_i}{n-1}\big) }}.
\label{phi-psi-sigma-n-expo}
\end{equation}
If $\psi^\sigma_{n}\neq 1$, then a \emph{constant forward Nash equilibrium} exists and is unique, with the constant optimal strategies $\pi^{i,*}$ given by
\begin{align}
\label{eq:optimal-control-n-player}
\pi^{i,*}_\cdot 
=
\frac1{{\nu_i^2+\sigma_i^2 \big(1+\frac{\theta_i}{n-1}\big) }}
\Big(\theta_i \sigma_i \Big(1+\frac1{n-1}\Big) \frac{\varphi^\sigma_n}{1-\psi^\sigma_n}  +  \mu_i \delta_i \Big).
\end{align}
The forward Nash equilibria is given by the $n$-pairs $\{(U^{i,*},\pi^{i,*})\}_{i=1,\cdots,n}$ where the $U^{i,*}$ is the solution of \eqref{eq:SPDE-step01} (see Example \ref{ex:ForwardExponentialUtility}) under the optimal constant strategies $\pi^{\cdot,*}$. 

The term $\lambda_i$ (see Example \ref{ex:ForwardExponentialUtility}), at equilibrium, is given by 
\begin{align}
\label{eq:lambda-n-player}
\nonumber
\lambda_i
& =
-\frac{\theta_i}{\delta_i} \Big(
\Big\{\frac n{n-1} \overline {\pi \mu} - \frac1{n-1} \pi^i \mu_i\Big\}
- \frac{\mu_i \sigma_i}{\nu_i^2+\sigma_i^2} 
\Big\{\frac n{n-1} \overline {\pi \sigma} - \frac1{n-1} \pi^i \sigma_i\Big\} \Big) 
\\ \nonumber
&
\qquad + \frac{\mu^2_i}{2(\nu_i^2+\sigma_i^2)}
  + \frac{\theta_i^2}{2\delta_i^2} \Big[  
	\Big\{\frac n{n-1} \overline {\pi \sigma} - \frac1{n-1} \pi^i \sigma_i\Big\}^2 \Big( \frac{ \sigma_i^2 }{\nu_i^2+\sigma_i^2} -1 \Big)
\\
&
\hspace{4.5cm}
- \Big\{\frac {n}{(n-1)^2} \overline {(\pi \nu)^2} - \frac1{(n-1)^2} (\pi^i \nu_i)^2\Big\} \Big],
\end{align}

where the relevant expressions for $\overline {\pi \sigma}$, $\overline {\pi \mu}$  and $\overline {(\pi \nu)^2}$ are given below in \eqref{eq:pisigma-nplayers}, \eqref{eq:pimu-nplayers} and \eqref{eq:pinu-nplayers}.
\end{theorem}

\begin{remark}
\label{rem:on-difference-with-Lacker}
We note that we do not solve the same problem studied at \cite{LackerZariphopoulou2017} but an equivalent one. However, imposing the scaling factor given by \cite[{Remark 2.5}]{LackerZariphopoulou2017} we recover the same results as in \cite[{Theorem 2.3}]{LackerZariphopoulou2017}.
\end{remark}

\begin{proof}
Injecting the condition $U_x/U_{xx}=-\delta_i$ in \eqref{eq:GeneralStrategySystem}, the system to be solved in order to ascertain the Nash equilibrium is, across $i\in\{1,\cdots,n\}$,
\begin{align*}
\pi^{i,*}_t (\nu_i^2+\sigma_i^2)
& 
=
\theta_i \sigma_i  \Big(\frac1{n-1}\sum_{k=1,k\neq i}^n \pi^{k,*}_t \sigma_{k} \Big)
+  \mu_i \delta_i
\\
&
=
\theta_i \sigma_i  \Big( \frac n{n-1} \overline{(\pi \sigma)}_t -\frac1{n-1} \pi^{i,*}\sigma_i\Big)
+  \mu_i \delta_i
\\
\Leftrightarrow \quad
&\pi^{i,*}_t 
=
\frac1{{\nu_i^2+\sigma_i^2 \big(1+\frac{\theta_i}{n-1}\big) }}
\Big(\theta_i \sigma_i \frac n{n-1} \overline{(\pi \sigma)}_t  +  \mu_i \delta_i \Big).
\end{align*}
The final line yields the expression for $\pi^{i,*}$ as a function of the unknown $\overline{\pi \sigma}$. To determine the latter, multiply both sides by $\sigma_i$ and average over $i\in\{1,\cdots,n\}$, this yields a solvability condition
\begin{align}
\label{eq:pisigma-nplayers}
\overline{(\pi \sigma)}_t
=
\overline{(\pi \sigma)}_t
\psi^\sigma_n
+
\varphi^\sigma_n
\ \ \Leftrightarrow\ \
\overline{\pi \sigma}
=\frac{\varphi^\sigma_n}{1-\psi^\sigma_n}\quad \textrm{ as long as }\quad \psi^\sigma_n\neq 1.
\end{align}
Plugging the expression $\overline{(\pi \sigma)}$ in that for $\pi^{i,*}$ yields the result. That the optimal strategies are constant is now obvious.

It remains to derive the expression for the $\lambda_i$'s. Just like for $\overline {\pi \sigma}$, we obtain an expression for $\overline {\pi \mu}$ by multiplying $\pi^{i,*}$ by $\mu_i$ and averaging on both sides, we have 
\begin{align}
\label{eq:pimu-nplayers}
\overline {\pi \mu} = \frac n {n-1}\cdot\frac{\varphi^\sigma_n}{1-\psi^\sigma_n}\cdot \psi^\mu_n + \phi^\mu_n
\quad\textrm{and}\quad
\overline{\pi \mu}^{(-i)} 
	= \frac n{n-1} \overline {\pi \mu} - \frac1{n-1} \pi^i \mu_i,
\end{align}
where we used \eqref{eq:identitiesBetweenAverages} and the quantities $\varphi^\mu_{n},\psi^\mu_{n}$ are defined as 
\begin{equation*}
\varphi^\mu_{n} 
:=\frac{1}{n}\sum_{k=1}^{n} \delta_{k} \frac{\mu^2_{k}}{\nu_{k}^{2}+\sigma_{k}^{2}(1+\frac{\theta_{k}}{n-1})}  \quad \text{ and } \quad 
\psi^\mu_n 
:= \frac{1}{n}\sum_{k=1}^{n}\theta_{k}\frac{\mu_k \sigma_{k}}{\nu_{k}^{2}+\sigma_{k}^{2}(1+\frac{\theta_{k}}{n-1})} .  
\end{equation*} 
Similarly, defining $\overline{(\pi \nu)^2}:=\frac1{n-1}\sum_{k\neq i} (\pi^k_t \nu_k)^2$ we have 
\begin{align}
\label{eq:pinu-nplayers}
\overline{(\pi \nu)^2}
=
\frac{1}{n} \sum_{i=1}^n 
\Big(\dfrac{\nu_i\theta_i \sigma_i \cdot \frac n{n-1} \cdot \frac{\varphi^\sigma_n}{1-\psi^\sigma_n}  +  \nu_i\mu_i \delta_i }{{\nu_i^2+\sigma_i^2 \big(1+\frac{\theta_i}{n-1}\big) }}\Big)^2.
\end{align}
Similarly to \eqref{eq:identitiesBetweenAverages}, we have 
$\overline{(\pi \nu)^2}^{(-i)}=\frac {n}{n-1} \overline {(\pi \nu)^2} - \frac1{n-1} (\pi^i \nu_i)^2$. Replacing these expressions in that for $\lambda_i$ in Example \ref{ex:ForwardExponentialUtility} the expression in the result's statement follows.
\end{proof}

From the forward utility machinery one can easily recover the classical case of utility optimization where one prescribes the utility map for the horizon time $T$ then proceeds to optimize.
\begin{example}[Recovering the classical utility problem from the forward one.]
If one would start the forward utility with (for some $0<T<\infty$)
\[
u^i_0(x):= - e^{-x/\delta_i -  T \lambda_i},
\]
then computations like those presented yield the forward utility map $U(x,t)$ as 
\[
U^i(x,t)=- e^{-x/\delta_i + (t-T)\lambda_i},\quad t\in[0,T]
\]
and in particular $U(x,T)=- e^{-x/\delta_i}$. In other words, our forward utility recovers as a particular case the classical exponential utility maximization problem (discussed in \cite{LackerZariphopoulou2017}).
\end{example}
\begin{corollary}[Single stock]
Let $\mu_i = \mu > 0,~\sigma_i = \sigma > 0$ and $\nu_i = 0$, for any $i = 1, \dots, n$. 
Defining constants as
\begin{equation}
\varphi^\sigma_{n} :=
\frac1n\sum_{i=1}^n   \frac{\delta_i}{1+\frac{\theta_i}{n-1}}
\ \ \text{ and } \ \ 
\psi^\sigma_n := 
\frac1{n-1}\sum_{i=1}^n   \frac{\theta_i}{1+\frac{\theta_i}{n-1} }.
\label{phi-psi-sigma-n-expo-single-stock}
\end{equation}
If $\psi^\sigma_{n}\neq 1$, then a \emph{constant forward Nash equilibrium} exists, with the constant optimal strategies $\pi^{i,*}$ given by
\begin{align*}
\pi^{i,*}_\cdot 
=
\frac{\mu}{{\sigma^2 \big(1+\frac{\theta}{n-1}\big) }}
\Big(\theta \Big(1+\frac1{n-1}\Big) \frac{\varphi^\sigma_n}{1-\psi^\sigma_n}  + \delta \Big).
\end{align*}
\end{corollary}

%
%
%
%
%

%
%
%
\section{The mean field game} 
\label{sec:MFG-forward}

By inspection of Theorem \ref{theo:nPlayerForwardNashGame} one sees that the optimal strategy and forward utility map for some agent depend on that agent's specific parameters (model parameters, initial wealth, risk tolerance and performance concern) and on certain averages of the parameters of all agents. This makes a case for a MFG approach to the game.

In this section and inspired by the results in the previous one, we formalize the concept of forward mean-field Nash game. We use the concept of \emph{type distributions} introduced in \cite{HuangMalhameCaines2006} and \cite{LackerZariphopoulou2017,lackersoret2020many}. We follow the construction presented in the latter.

We focus on initial forward utilities at time $t=0$ that are of exponential type, 
\begin{align*}
U^i(x,m,0) & = -\exp\Big\{ -\frac1{\delta_i} (x -\theta_i m)\Big\},
\end{align*}
where we refer to the constants $\delta_i>0$ and $\theta_i\in[0,1]$ as \emph{personal risk tolerance} and \emph{competition weight} parameters, respectively.

For the $n$-agent game, we define for each agent $i=1,\ldots,n$ the \emph{type vector}
\begin{equation*}
\zeta_i := (x^i_0,\delta_{i},\theta_{i},\mu_{i},\nu_{i},\sigma_{i}),
\end{equation*}
which characterizes perfectly each agent $i$. 
These \textit{type vectors} induce an empirical measure, called the \emph{type distribution}, which is the probability measure on the \emph{type space}
\begin{align}
\label{def:exp-type-space}
\cZ^e  := \bR \times (0,\infty) \times [0,1] \times (0,\infty) \times [0,\infty) \times [0,\infty), 
\end{align}
given by
\[
m_n(A) = \frac{1}{n}\sum_{i=1}^n1_A(\zeta_i), \ \text{ for Borel sets } A \subset \cZ^e .
\]
Assume now that as the number of agents becomes large, $n\rightarrow\infty$, the above empirical measure $m_n$ has a weak limit $m$, in the sense that $\int_{\cZ^e } f\,dm_n \rightarrow \int_{\cZ^e } f\,dm$ for every bounded continuous function $f$ on $\cZ^e $. For example, this holds almost surely if the $\zeta_i$'s are i.i.d.\ samples from $m$. Let $\zeta=(\xi,\delta,\theta,\mu,\nu,\sigma)$ denote an $\cZ^e$-valued random variable with this limiting distribution $m$.

The \emph{mean field game} (MFG) defined next allows us to derive the limiting strategy as the outcome of a self-contained equilibrium problem, which intuitively represents a game with a continuum of agents with type distribution $m$. Rather than directly modelling a continuum of agents, we follow the MFG paradigm of modelling a single \textit{generic agent}, who we view as randomly selected from the population. The probability measure $m$ represents the distribution of type parameters among the continuum of agents; equivalently, the generic agent's type vector is a random variable with law $m$. Heuristically, each agent in the continuum trades in a single stock driven by two Brownian motions, one of which is unique to this agent and one of which is common to all agents. We extend the Forward Nash equilibrium of Definition \ref{def:ForwardNashEquilibrium} to the MFG setting below.

\subsection{Agents through type-distribution and the market}
\label{sec4.1}

Let $(\Omega, \cF,\bF = (\cF)_{t \geqs 0},\bP)$ be a stochastic basis supporting two independent Brownian motions $W = (W_t)_{t \geqs 0}$ and $B = (B_t)_{t \geqs 0}$ together with a random vector $\zeta$ having distribution $m$ and given by 
\[
\zeta = (\xi,\delta ,\theta ,\mu ,\nu ,\sigma),
\]
with values in the space $\cZ^e $ defined in \eqref{def:exp-type-space} and independent of $W$ and $B$. Let $\bF= (\cF_t)_{t \in [0,T]}$ denote the smallest filtration satisfying the usual assumptions for which $\zeta$ is $\cF_0$-measurable and both $W$ and $B$ are adapted. Let also $\bF^B=(\cF^B_t)_{t \in [0,T]}$ denote the natural filtration generated by the Brownian motion $B$.

The \emph{generic agent's} wealth process solves
\begin{align}
\label{def:X-MFG}
dX_t = \pi_t(\mu dt + \nu dW_t + \sigma dB_t), \quad X_0 = \xi, 
\end{align}
where the portfolio strategy must belong to the admissible set $\cA_{\mathrm{MF}}$ of self-financing $\bF$-progressively measurable real-valued processes $(\pi_t)_{t\geqs 0}$ satisfying the square-integrability condition $\bE[\int_0^T|\pi_t|^2dt] < \infty$ for any $T\in [0,\infty)$.
The generic agent's initial wealth is given by $\xi$, whereas $(\mu,\nu,\sigma)$ are the market parameters. In the sequel, the parameters $\delta$ and $\theta$ will affect the risk preferences of the generic agent. Each agent among the continuum will have different preference parameters and hence these six parameters are $\cF_0$-random, and each has the exact same interpretation an in the $n$-player game of the earlier section.

\subsection{The equilibrium}

The formulation of the forward Nash game of Section \ref{sec:FRPCwithaverage} drives the formulation of the Mean-field game we discuss here. Recall that in the MFG-formulation the \emph{generic agent} has no influence on the average wealth of the continuum of agents, as but one agent amid a continuum of agents. We next introduce the concept of the main object of interest the \emph{MF-Forward relative performance equilibrium}.

We introduce the regularity requirements for the utility.

\begin{assumption}
\label{ass:regularity-u-mean-field}
Assume that the derivatives $U_t(x,t)$, $U_{x}(x,t)$ and $U_{xx}(x,t)$ exists for $t\geqs 0,~x\in\bR$, $\bP$-a.s.
\end{assumption}

As in Section \ref{sec:FRPCwithaverage-n-1}, Assumption \ref{ass:regularity-u-mean-field} implies the It\^o decomposition of map $U$
\begin{align*}
    d U(x,t) = U_t(x,t) dt.
\end{align*}
Given this market setup we next define our concept of equilibrium.

\begin{definition}[MF-Forward CARA relative performance equilibrium (for the generic manager)]
\label{def:MFG-Forward-problem}
Let $(\overline{X}_t)_{t\geqs 0}$ be the $\bF^{B}$-adapted square integrable stochastic process representing the average wealth of the continuum of agents. Let $\pi\in \cA^{\mathrm{MF}}$ and $X^\pi$ solve \eqref{def:X-MFG} with $\pi$. 

The $\bF^{\mathrm{MF}}$-progressively measurable random field $(U(x,t))_{t\geqs 0}$  is an \emph{MF-forward relative performance} for the \emph{generic manager} if, for all $t\geqs 0$, the following conditions hold:
\begin{enumerate}[i)]
	\item The mapping $x \mapsto U(x,t)$, is $\bP$-a.s.~strictly increasing and strictly concave;
	\item  For any $\pi \in \cA^{\textrm{MF}}$, $U( X_t^\pi-\theta \overline X_t,t)$ is a (local) supermartingale and $X^\pi$ is the \emph{generic agent's} wealth process solving \eqref{def:X-MFG} for the strategy $\pi$;
	\item There exists $\pi^{*}\in \cA^{\textrm{MF}}$ such that $U( X^*_t-\theta \overline X_t,t)$ is a (local) martingale where $X^*$ solves \eqref{def:X-MFG} with $\pi^{*}$ plugged in as the strategy;
	\item We call $\pi^*$ of point iii) a \emph{MF-equilibrium} if $\overline X_t = \bE[X^*_t| \cF^B_t]$ for all $t\geqs 0$ where where $X^*$ solves \eqref{def:X-MFG} with $\pi^{*}$ plugged in as the strategy.
\end{enumerate}
We denote the triplet $(U,\pi^*,\overline X,)$ satisfying i)-iv) the \emph{MF-Forward relative performance equilibrium}. 
 An MF-equilibrium is \emph{constant} if there exists an $\cF_0^{\textrm{\textrm{MF}}}$-measurable RV $\pi^*$ such that $\pi_t=\pi^*$, for all $t\geqs 0$.
\end{definition}

The last point can be understood as a fixed point argument which creates a compatibility condition between the generic agent within the continuum of agents. In fact, conditionally on the BM $B$ each agent faces an independent noise $W$ and an independent type vector $\zeta$. As in Mean-field games \cite{LackerZariphopoulou2017,lackersoret2020many}, conditionally on $B$, all agents faces i.i.d. copies of the same optimization problem. The law of large numbers suggests that the average terminal wealth of the whole population should be $\bE[X_t^*|\cF^B_t]$.

Our construction allows us to identify $\bE[X_t^*|\cF^B_t]$ with a certain dynamics and, in turn, treat this component as an additional uncontrolled state process. This avoids altogether the conceptualization of the master equation for models with different types of agents. The latter is left for future research.

\subsection{Solving the optimization problem}

We now present the main result of this section which is the existence of a \emph{MF-Forward CARA relative performance equilibrium} for the generic manager according to Definition \ref{def:MFG-Forward-problem} within the context of time-monotone forward utilities.

From the methodological point of view, the problem is solved as before. Apply It\^o formula to $U(Z^\pi_t,t)$, determine the optimal strategy $\pi^*$ and the consistency condition (the PDE) for $U$ such that the first three conditions of Definition \ref{def:MFG-Forward-problem} hold. The last condition, to show that $\pi^*$ is indeed the MFG Forward equilibrium follows by construction as we will see.

\begin{theorem}
\label{theo:MFG-solution}
Take a generic agent $\zeta = (\xi,\delta ,\theta ,\mu ,\nu ,\sigma)$ and assume that $\delta>0$, $\theta\in[0,1]$, $\mu>0$, $\sigma\geqslant 0$, $\nu\geqslant 0$ such that $\sigma^2+\nu^2>0$.

Assume the following constants are finite
\begin{align*}
\psi^\sigma:= &\bE\Big[\theta\frac{ \sigma^2}{\nu^2+\sigma^2}\Big],
\quad
\varphi^\sigma:= \bE\Big[\delta \frac{\mu\sigma }{\nu^2+\sigma^2}\Big],
\\
\psi^\mu:=&\bE\Big[\theta\frac{ \mu \sigma}{\nu^2+\sigma^2}\Big],
\quad \textrm{and}\quad 
\varphi^\mu:=\bE\Big[ \delta \frac{\mu^2 }{\nu^2+\sigma^2} \Big].
\end{align*}
Assume that $\psi^\sigma\neq 1$. Then there exists a unique constant MF-Forward CARA relative performance equilibrium in the sense of Definition \ref{def:MFG-Forward-problem}. 

The constant MF-equilibrium strategy is unique and is given by 
\begin{align}
\label{eq:optimstrategy-MFG}
\pi^* 
=
\frac1{{\nu^2+\sigma^2}}\Big( \theta \sigma \frac{\varphi^\sigma}{1-\psi^\sigma}  +  \mu \delta \Big),
\end{align}
constrained to the identity
\begin{align*}
    \bE[{\sigma \pi^*}]=\frac{\varphi^\sigma}{1-\psi^\sigma}<\infty.
\end{align*}
The MF-forward CARA relative performance utility map under Assumption  \ref{ass:regularity-u-mean-field} is the unique solution of the PDE with stochastic coefficients
\begin{align}
\label{eq:MF-SPDE}
\nonumber
U_t 
&=
\theta \Big( 
\frac{\varphi^\sigma}{1- \psi^\sigma}\cdot 
\psi^\mu+\varphi^\mu
- \mu \frac{ \sigma }{\nu^2+\sigma^2} \cdot\frac{\varphi^\sigma}{1-\psi^\sigma}
  \Big) U_x
\\
&
\qquad
+  \frac{\mu^2}{2(\nu^2+\sigma^2)}   \frac{(U_x)^2}{U_{xx}}
+ \frac 12 U_{xx} \cdot \theta^2 
\Big(\frac{\varphi^\sigma}{1-\psi^\sigma}\Big)^2
\Big(\frac{\sigma^2}{\nu^2+\sigma^2}-1 \Big).
\end{align}
When the initial condition is $U(x,0)=u_0(x)=-e^{-x/\delta}$, i.e.~the exponential preferences, $U$ is given explicitly by $U(x,t)=u_0(x)e^{t \lambda}$ with $\lambda$ given by 
\begin{align}
\label{lambdaMFG-inMainTheo}
\lambda= -\frac{ \theta}{\delta} \overline{ \mu \pi}
+ \frac{1}{2(\nu^2+\sigma^2)}\Big( \mu + \frac{\theta}{\delta} \sigma \overline{ \sigma \pi} \Big)^2
- \frac{\theta^2}{2\delta^2}\Big(\overline{ \sigma \pi}\Big)^2,
\end{align}
where $\overline{ \sigma \pi}$ and $\overline{ \mu \pi}$ are given by \eqref{eq:ConsistencyExpressions-alphasigma} and \eqref{eq:ConsistencyExpressions-alphanu} respectively.
If $\psi^\sigma=1$, then there exists no constant MF-equilibrium.
\end{theorem}
By comparing the statements of Theorem \ref{theo:nPlayerForwardNashGame} and 
Theorem \ref{theo:MFG-solution}
(and same happens for the respective Single (common) Stock Corollaries) one easily sees that as $n\to\infty$ the strategies, weights ($\phi^\cdot_n$ and $\psi^\cdot_n$) and forward-utility map in Theorem \ref{theo:nPlayerForwardNashGame} converge to the respective quantities appearing in Theorem \ref{theo:MFG-solution}.
\begin{remark}
We point out that the interaction of the generic agent with the continuum is only performed through the common noise $B$. That can be seen by the term $ \frac{n}{(n-1)^2} \overline {(\pi \nu)^2} - \frac1{(n-1)^2} (\pi^i \nu_i)^2$ from $\lambda_i$'s in \eqref{eq:lambda-n-player} converging to zero as $n \to \infty$, as we have by \eqref{eq:lambda for the MFG} (compare with \eqref{lambdaMFG-inMainTheo}). We can interpret it via the standard mean-field approximation, the individual's impact on the others is negligible for the infinite system.
\end{remark}
\begin{remark}
\label{rem:about-MF-result-and-Lacker}
In contrast with Remark \ref{rem:on-difference-with-Lacker}, here we recover the result from \cite[Theorem 2.10]{LackerZariphopoulou2017} as the scaling factors converge to $1$ (as $n \to \infty$).
Hence, due to space constraints we defer the reader to \cite[Section 2.3]{LackerZariphopoulou2017} for the discussion of the equilibria. 
\end{remark}
\begin{proof}
We proceed in several steps in order to construct the constant MF-equilibrium. To that end we must solve ii)-iii) in Definition \ref{def:MFG-Forward-problem} for a given $\overline X$ process associated to $\pi\in \cA_{\textrm{MF}}$. Condition iv), for MF-equilibrium allows us to focus only on processes of the form $\overline X_t = \bE[X^\pi_t| \cF^B_t]$ where $X^\pi$ solves \eqref{def:X-MFG} for a constant strategy $\pi$ (i.e.~$\cF^{\textrm{MF}}_0$-measurable) satisfying $\bE[\pi^2]<\infty$.

\emph{Step 0. The dynamics of the average wealth process.} To solve the above problem given $(\overline X_t)_{t\geqs 0}$ it suffices to restrict ourselves to processes $(\overline X_t)_{t\geqslant 0}$ satisfying $\overline X_t = \bE[X^\pi_t| \cF^B_t]$ $\bP$-a.s..~We then have 
\begin{align}
\label{eq:Dynamics for Average Wealth}
\nonumber
\overline X_t 
= \bE[X^\pi_t| \cF^B_t]
&
= \bE\Big[ \xi + \int_0^t \mu \pi  ds + \int_0^t\nu \pi dW_s + \int_0^t\sigma \pi dB_s \Big| \cF^B_t\Big]
\\
&
= \bar \xi + \int_0^t \overline{\mu \pi_s}  ds + \int_0^t \overline{\sigma \pi_s} dB_s,
\end{align}
where, for consistency of notation with the previous section, we denote 
\begin{align*}
\bar \xi:=\bE[ \xi],\quad
\overline{\mu \pi}:=\bE[\mu \pi] 
\quad\textrm{and}\quad
\overline{\sigma \pi}:=\bE[ \sigma \pi].
\end{align*}
Hence for $\pi \in \cA^{\textrm{MF}}$ and as in the previous section we can define the dynamics of the process $Z^\pi = X^\pi - \theta \overline X$ 
\begin{align*}
d Z^\pi_t 
&= \big(\mu \pi_t-\theta \overline{\mu \pi} \big)dt + \nu \pi_t d W_t +\big(\sigma \pi_t-\theta \overline{\sigma \pi} \big)dB_t,
\quad Z^\pi_0=\xi - \theta \overline \xi,
\end{align*} 
and solve the MFG Forward utility problem in Definition \ref{def:MFG-Forward-problem} with its help. 

Hence applying It\^o's formula to $U(Z^\pi_t,t)$ yields
\begin{align}
\label{eq:generalSDEafterItoWentzell-MFG}
\nonumber
 d U(Z^\pi_t ,t) 
&= U_t(Z^\pi_t ,t)dt + U_x(Z^\pi_t ,t) d Z^\pi_t + \frac12 U_{xx}(Z^\pi_t ,t) d \langle Z^\pi_t \rangle 
\\ \nonumber
& 
=\Big[U_t(Z^\pi_t ,t)   
+ U_x(Z^\pi_t ,t) \big(\mu \pi_t - \theta \overline{\mu \pi} \big)\\
&
\qquad
+\frac12 U_{xx}(Z^\pi_t ,t) \Big( (\nu \pi_t )^2 + \big(\sigma \pi_t - \theta \overline{ \sigma \pi } \big)^2\Big)\Big]dt,
\\ \nonumber
&
\qquad
+ U_x(Z^\pi_t ,t) \nu\pi_t dW_{t} 
+ U_x(Z^\pi_t ,t)\big(\sigma \pi_t- \theta \overline{ \sigma \pi} \big)dB_t,
\end{align}
with $U(Z^\pi_0 ,0)=U(\xi-\theta \overline \xi,0)=-\exp\{-(\xi-\theta \overline \xi)/\delta \}$ and we used that the $B,W$ are all i.i.d. Exact calculations on deriving \eqref{eq:generalSDEafterItoWentzell-MFG} are presented in the Section \ref{sec:appendix}.

\emph{Step 1. Finding the candidate optimal strategy $\pi^*$.} As before, the process $U(Z^\pi_t,t)$ becomes a Martingale at the optimum $\pi$. Direct computations using first order conditions ($\partial_{\pi} \textrm{``drift''}=0$) yield
\begin{align}
\label{eq:optimstrategy-MFG-vAUXILIARY}
\nonumber
 0 &+ U_x \cdot \big( \mu -0 \big) + \frac12 U_{xx} \Big[2 \pi \nu^2 + 2\big(\sigma \pi_t- \theta \overline{\sigma \pi} \big) \sigma \Big] =0
\\
&
\Rightarrow \qquad \pi^*_t ({\nu^2+\sigma^2})
=
\theta \sigma \overline{\sigma\pi} -  \mu \frac{U_x(Z^\pi_t ,t)}{U_{xx}(Z^\pi_t ,t)}
=
\theta \sigma \overline{\sigma\pi} +  \mu \delta,  
\end{align} 
where we injected the CARA constraint $U_x/U_{xx}=-\delta$, for all $t$. By inspection it is clear that $\pi^*$ is a $\cF_0^{\textrm{MF}}$-measurable RV which is independent of time and is well-defined as long as $\overline{\sigma\pi}$ is finite.

\emph{Step 2. The optimality of the strategy.} The argument is similar to that in \cite{LackerZariphopoulou2017}. The original constant strategy $\pi$ if a MF-equilibrium if and only if for all $t\geqslant 0$
\begin{align*}
\bE[X^{\pi}_t| \cF^B_t]
&=\bE[X^{\pi^*}_t| \cF^B_t]\quad a.s.
\\
& 
\Leftrightarrow \quad 
\bar \xi + \overline{\mu \pi}\, t + \overline{\sigma \pi} B_t
=
\bar \xi + \overline{\mu \pi^*}\, t + \overline{\sigma \pi^*} B_t\quad a.s.
\end{align*}
Taking expectations on both sides implies that $\pi$ is a MG-equilibrium if and only if the following two conditions holds
\[
\overline{\mu \pi}=\overline{\mu \pi^*}
\quad\textrm{and}\quad
\overline{\sigma \pi}=\overline{\sigma \pi^*}.
\]
Using \eqref{eq:optimstrategy-MFG-vAUXILIARY}  with $U_x/U_{xx}=-\delta$ and the expressions for $\varphi^\sigma,\psi^\sigma$ one derives that
\begin{align*}
\sigma \pi^* & =  \theta\frac{ \sigma^2}{\nu^2+\sigma^2} \overline{\sigma \pi} +  \delta \frac{\mu\sigma }{\nu^2+\sigma^2}
\quad  \Rightarrow  \quad 
\overline{\sigma \pi^*}
= \overline{\sigma \pi} \psi^\sigma + \varphi^\sigma,
\end{align*} 
using that $\overline{\sigma \pi}=\overline{\sigma \pi^*}$ yields solvability if $\psi^\sigma=\bE\big[\theta\frac{ \sigma^2}{\nu^2+\sigma^2}\big]\neq 1$. The same procedure deals with the condition $\overline{\mu \pi}=\overline{\mu \pi^*}$. We then have 
\begin{align}
\label{eq:ConsistencyExpressions-alphasigma}
\overline{\sigma \pi^*}&=\overline{\sigma \pi}=\frac{\varphi^\sigma}{1-\psi^\sigma}=\textrm{Const},\\
\label{eq:ConsistencyExpressions-alphanu}
\overline{\mu \pi^*} &=\overline{\mu \pi} 
=  \frac{\varphi^\sigma}{1- \psi^\sigma}\cdot 
\psi^\mu+\varphi^\mu 
=\textrm{Const.}
\end{align}
Injecting these identities in the expression for $\pi^*$ we find \eqref{eq:optimstrategy-MFG}. 

For the non-solvability statement, if the equation \eqref{eq:ConsistencyExpressions-alphanu} has $\psi^\sigma=1$ and $\varphi^\sigma\neq 0$ then the equation has no solution and hence no constant MF-equilibrium exists.  The case  $\psi^\sigma=1$ and $\varphi^\sigma= 0$ is impossible. Since $\mu>0$ and $\delta>0$ by assumption, it implies that $\sigma=0$ and hence that $\psi^\sigma=0 $ contradicting the condition $\psi^\sigma=1$.

\emph{Step 3. Finding the consistency PDE and the Utility map.}  We do not carry out this step explicitly, nonetheless, injecting the expression of $\pi^*$, $\overline{\sigma \pi}$ and $\overline{\mu \pi}$ in the drift term of \eqref{eq:generalSDEafterItoWentzell-MFG} and simplifying, we find the necessary equation \eqref{eq:MF-SPDE}, i.e. the consistency condition the random field $U$ must satisfy to that the required properties in Definition \ref{def:MFG-Forward-problem} hold.

Just like in Example \ref{ex:ForwardExponentialUtility}, the time-monotone forward utility  equation \eqref{eq:MF-SPDE} can be solved and indeed one has a simplified version. We have
\begin{align}
 \label{eq:GenericAgentCARAForwardUtility} 
U(x,t) & = -e^{-{x}/{\delta}+ t \lambda},
\end{align}
where the $\cF_0^{\textrm{MF}}$-measurable RV $\lambda$ is given by (using \eqref{eq:ConsistencyExpressions-alphasigma} and \eqref{eq:ConsistencyExpressions-alphanu})
\begin{align}
\label{eq:lambda for the MFG}
\lambda 
&= -\frac{ \theta}{\delta} \overline{ \mu \pi}
+ \frac{1}{2(\nu^2+\sigma^2)}\Big( \mu + \frac{\theta}{\delta} \sigma \overline{ \sigma \pi} \Big)^2
- \frac{\theta^2}{2\delta^2}\Big(\overline{ \sigma \pi}\Big)^2 
\\ \nonumber
&=
-\frac{\theta}{\delta} \Big( 
\frac{\varphi^\sigma}{1- \psi^\sigma}\cdot \psi^\mu+\varphi^\mu
- \mu \frac{ \sigma }{\nu^2+\sigma^2} \cdot\frac{\varphi^\sigma}{1-\psi^\sigma}\Big)\\
\nonumber
&\qquad+  \frac{\mu^2}{2(\nu^2+\sigma^2)}  
+ \frac{\theta^2}{2\delta^2} 
\Big(\frac{\varphi^\sigma}{1-\psi^\sigma}\Big)^2
\Big(\frac{\sigma^2}{\nu^2+\sigma^2}-1 \Big).
\end{align}


\emph{Step 4. The MFG forward utility dynamics.} Injecting the consistency PDE \eqref{eq:MF-SPDE} in the expression for  $d U(Z^\pi_t ,t)$ given in \eqref{eq:generalSDEafterItoWentzell-MFG} yields,  
\begin{align*}
 d U(Z^\pi_t ,t) 
&= 
\frac12 \frac{U_{xx}(Z^\pi_t ,t) }{(\nu^2+\sigma^2)}  \Big| \pi_t (\nu^2+\sigma^2)
-\Big( \theta \sigma \cdot \frac{\varphi^\sigma}{1- \psi^\sigma} + \mu \delta \Big)   \Big|^2dt
\\
&
\qquad
+ U_x(Z^\pi_t ,t) \nu\pi_t dW_{t} 
+ U_x(Z^\pi_t ,t)\Big(\sigma \pi_t- \theta\cdot \frac{\varphi^\sigma}{1- \psi^\sigma} \Big)dB_t.
\end{align*}
\end{proof}

We close with a corollary regarding the common stock case.
\begin{corollary}[Single stock]
Let $\mu,\sigma,\nu$ be deterministic with $\nu  = 0, \mu,\sigma > 0$. 
Defining constants as
\begin{equation}
\varphi :=
\bE[\delta]
\ \ \text{ and } \ \ 
\psi := \bE[\theta].
\label{phi-psi-sigma-mean-field-single-stock}
\end{equation}
Then, if $\psi\neq 1$ then a \emph{constant MF-equilibrium} exists, with the constant optimal strategy $\pi^{*}$ given by
\begin{align*}
\pi^{*}_\cdot 
=
\frac{\mu}{\sigma^2}
\Big(\theta \frac{\varphi}{1-\psi}  + \delta \Big).
\end{align*}
\end{corollary}

%
%
%
%
%

%
%

\subsection{Mean-field dynamic model selection with large horizons}
\label{sec:dynamicselection}


Over the time interval $[0,\infty)$ our generic agent selects a sequence of horizon time $(T_j)_{j\in\bN_0}$ (such that $T_0=0$, $T_{j+1}-T_j>0$ and $\lim_j T_j=\infty$) on which the agent assesses and updates the  market model by adjusting the model's coefficients. Comparing with \eqref{def:X-MFG} the agent models the stock as 
\begin{align}
\label{stock-rolling}
\frac{dS_{t}^{j}}{S_{t}^{j}} 
& 
=\mu_{j}dt+\nu _{j}dW_{t}+\sigma_{j}dB_{t},\quad S_{T_j}=s_{j}, \:\:
t\in[T_j,T_{j+1}],
\end{align}
where the index $j$ represents the model specification at time $T_j$. The associated wealth process of the \emph{generic agent} is
\begin{align*}
dX^j_t = \pi_t(\mu_j dt + \nu_j dW_t + \sigma_j dB_t), \quad X_{T_j} = \xi_j, \:\: t\in[T_j,T_{j+1}].
\end{align*}
Following the earlier constructions of this section, assume that at time $T_0=0$ the agent starts with initial utility $u_0(x)=-e^{-x/\delta}$. Then using the results of Theorem \ref{theo:MFG-solution}, the agent's forward utility map is given by 
\begin{align*}
U(x,t)= -e^{x/\delta} e^{t \lambda_0}=u_0(x)e^{t \lambda_0},\quad t\in[T_0,T_1]=[0,T_1],
\end{align*}
where $\lambda_0$ is the version of \eqref{eq:lambda for the MFG} for the type of the agent over the time interval $[T_0,T_1]$ and all the coefficients correspond to a type $\zeta_0$, i.e. $\lambda(\zeta_0)=\lambda_0$, with
\begin{align}
\label{eq:lambda as map of zeta}
\lambda_0=\lambda(\zeta_0)
:= -\frac{ \theta}{\delta} \overline{ \mu \pi}
+ \frac{1}{2(\nu^2+\sigma^2)}\Big( \mu + \frac{\theta}{\delta} \sigma \overline{ \sigma \pi} \Big)^2
- \frac{\theta^2}{2\delta^2}\Big(\overline{ \sigma \pi}\Big)^2. 
\end{align}
At time $T_1$, the generic agent assesses the previous model specification and chooses new coefficients (leading to a change in type, say from $\zeta_0$ to $\zeta_1$). The agent then carries out the optimization program over $t\in[T_1,T_2]$ but starting from initial utility $U(x,T_1)$. Under the assumption of constant coefficients Theorem \ref{theo:MFG-solution}, yields, 
\begin{align*}
U(x,t)= \Big(u_0(x)e^{T_1 \lambda_0}\Big) e^{(t-T_1)\lambda_1},\quad t\in[T_1,T_2],
\end{align*}
where $\lambda_1=\lambda(\zeta_1)$ (given by \eqref{eq:lambda as map of zeta}) depends only on information at time $T_1$. 
Quick calculations generalize to any time horizon $T_j$. Assume we work on the time interval $[T_j,T_{j+1}]$. Stemming from previous calculations, it is easy to see that the initial condition for the forward utility problem is
\begin{align*}
 U(x,T_j)=u_0(x) \prod_{k=1}^j e^{(T_k-T_{k-1})\lambda_{k-1}}
\end{align*}
(with the convention that if $j<1$ then $\prod_{k=1}^j \dots=0$) and the MFG forward utility is for all $t\in[T_j,T_{j+1}],~ j>1$ and using that $\lambda_j=\lambda(\zeta_j)$.
\begin{align*}
U(x,t)&
= U (x,T_j) e^{(t-T_{j})\lambda_{j}}
= u_0(x) \prod_{k=1}^j e^{(T_k-T_{k-1})\lambda_{k-1}} \cdot e^{(t-T_{j})\lambda_{j}},
\\
&
= u_0(x) \exp\Big\{{T_{1}(\lambda_{0}-\lambda_{1})} + {T_2(\lambda_{1}-\lambda_{2})} + \cdots+ {T_{j}(\lambda_{j-1}-\lambda_{j})}\Big\} e^{t \lambda_{j}}.
\end{align*}

There are two points to highlight. Firstly, the agent needs to carry information of what happened in the past in order to have time-consistency at present time.
Secondly, this construction also allows the agents to change not just the model specification $(\mu,\nu,\sigma)$ but also their type including risk parameter $\delta$ and performance-concern level $\theta$. The initial wealth is fixed from the previous time interval.

%
%
%

\section{Outlook and open questions}
\label{sec:Outlook}

In this work we considered two optimal portfolio management problems under forward utility performance concerns. We presented a simplified setting allowing for explicit calculations of the optimal control value function, strategies and an intuitive validation that the finite-play game reaches the mean-field game in the limit. 

This work provides a proof-of-concept for the forward mean-field utility construction leaving open many questions. Generalizing the dynamics of the forward utility \eqref{eq:SDEforForwardUtilityField} to a fully It\^o-dynamics and stochastic strategies is also open. A crucial tool for such would be a general It\^o-Wentzell-Lions chain rule as developed in \cite{platonov2019ito}. Such an approach would require \cite{Zitkovic2009}, \cite{KarouiMrad2013}.

Here we addressed only the exponential-utilities (CARA) and left the power-case (CRRA) open. Even within  \eqref{eq:SDEforForwardUtilityField}, one can build towards the CRRA case in \cite{LackerZariphopoulou2017} or include the consumption problem \cite{lackersoret2020many}; for the general forward utility case see \cite{KarouiHillairetMrad2018}. Also open is the so-called mean-field aggregation problem where different agents use utility maps from different families, e.g.~CRRA and CARA: \cite{KarouiHillairetMrad2019} would be a starting point for the finite-player case while the mean-field case would requires the multi-class approach of \cite[Section 8]{BensoussanFrehseYam2013} with the parameterization technique of from our Section \ref{sec:MFG-forward}. Many other questions can be posed in this context of mean-field forward utilities, ranging from possible non-solvability \cite{FreiDosReis2011}, to risk-sharing \cite{BielagkLionnetDosReis2017}, ergodic problems \cite{GechunZariphopoulou2016ergodic} and associated numerics \cite{GobetMrad2018}.



\bibliographystyle{abbrv}


\appendix
\section{Supplementary calculations}
\label{sec:appendix}


\begin{proof}[of Proposition \ref{prop:BestResponses-n-playerGame-average-n-1}]
We recall the optimal strategy is given by  \eqref{eq:optimstrategy},
where we define 
\begin{align*}
\widehat \sigma:= \overline{(\pi \sigma)}^{(-i)}_t, 
\ \ 
B^\nu_t:=\theta_i^2 \frac1{(n-1)^2}\sum_{k\neq i} (\pi^k_t \nu_k)^2,
\ \ 
M^\mu_t:=\theta_i\overline{(\pi \mu)}^{(-i)}_t= \frac{\theta_i }{n-1}\sum_{k\neq i} \pi_t^k \mu_{k}.
\end{align*}
The drift of \eqref{eq:generalSDEafterItoWentzell} becomes (we omit the argument in $U_t,U_x,U_{xx}$ and use $\widehat \sigma:= \overline{(\pi \sigma)}^{(-i)}_t$)
\begin{align*}
U^i_t+ U^i_x  &\big(\pi_t^i \mu _{i}- M^\mu_t \big)
+\frac12 U^i_{xx}\Big[ (\pi^i_t \nu_i)^2 + B^\nu_t +  \big(\pi_t^i\sigma _{i}- \theta_i\overline{(\pi \sigma)}^{(-i)}_t \big)^2\Big]
\\
= & 
\Big( U^i_t - M^\mu_t U^i_x + \frac12 U^i_{xx}B^\nu_t  \Big)
+\frac12 U^i_{xx}\Big[ \big(\theta_i \widehat \sigma \big)^2-(\pi_t^i)^2 (\nu_i^2 + \sigma _{i}^2)  \Big]
\\
= & 
U^i_t 
+ {U^i_x} \Big[ {\theta_i \sigma_i \widehat \sigma} \mu_i  \frac1{\nu_i^2+\sigma_i^2} - M^\mu_t \Big] 
-\frac{\mu_i^2}2 \frac1{\nu_i^2+\sigma_i^2} \frac{ ( U^i_x)^2 }{U^i_{xx}}
\\
& \quad 
+\frac12 U^i_{xx}
\Big\{ B^\nu_t + \big(\theta_i \widehat \sigma \big)^2 -\frac1{\nu_i^2+\sigma_i^2}\big({\theta_i \sigma_i \widehat \sigma}\big)^2\Big\}
\\
= & 
U^i_t 
+ {U^i_x} \Big[ \frac{ \mu_i \theta_i \sigma_i \widehat \sigma }{\nu_i^2+\sigma_i^2} - \theta_i\overline{(\pi \mu)}^{(-i)}_t \Big] 
-\frac{\mu_i^2}{2(\nu_i^2+\sigma_i^2)} \frac{ ( U^i_x)^2 }{U^i_{xx}}
\\
& \quad 
+\frac12 U^i_{xx}
\Big\{ \theta_i^2 \frac1{(n-1)^2}\sum_{k\neq i} (\pi^k_t \nu_k)^2 
+ \big(\theta_i \widehat \sigma \big)^2 \Big[1-\frac{\sigma_i^2}{\nu_i^2+\sigma_i^2}\Big] \Big\}.
\end{align*}

Equation \eqref{eq:SPDE-step01} now follows as $U^i_t$ needs to be chosen such that the equation is zero.
We inject in the drift of \eqref{eq:generalSDEafterItoWentzell} the expression  \eqref{eq:SPDE-step01} and obtain a simplified version 
\begin{align*}
&
-\Big\{
{U^i_x} \Big[ {\theta_i \sigma_i \widehat \sigma} \mu_i  \frac1{\nu_i^2+\sigma_i^2} - M^\mu_t \Big] 
-\frac{\mu_i^2}2 \frac1{\nu_i^2+\sigma_i^2} \frac{ ( U^i_x)^2 }{U^i_{xx}}+\frac12 U^i_{xx}
\Big\{ B^\nu_t + \big(\theta_i \widehat \sigma \big)^2\\
&-\frac1{\nu_i^2+\sigma_i^2}\big({\theta_i \sigma_i \widehat \sigma}\big)^2\Big\}\Big\} 
+
U^i_x \big(\pi_t^i \mu _{i}- M^\mu_t \big)
\\
&
+\frac12 U^i_{xx}\Big[ (\pi^i_t \nu_i)^2 + B^\nu_t
+  \big(\pi_t^i\sigma _{i}\big)^2  - 2 \pi_t^i\sigma _{i} \theta_i \widehat \sigma  + \big(\theta_i \widehat \sigma \big)^2 \Big]
\\
= & 
\frac{ U^i_{xx}}2 \frac1{\nu_i^2+\sigma_{i}^2}\Big( \big(\pi_t^i)^2(\nu_i^2+\sigma_{i}^2)^2  - 2 \big({\pi_t^i}(\nu_i^2+\sigma_{i}^2)\big) \Big(\sigma _{i} \theta_i \widehat \sigma - \mu _{i} \frac{U^i_x}{U^i_{xx}}\Big) \Big)
\\
&
+
- \frac1{\nu_i^2+\sigma_i^2} \frac{U^i_{xx}}{2}\frac2{U^i_{xx}} 
\Big\{
{U^i_x} \Big[ {\theta_i \sigma_i \widehat \sigma} \mu_i     \Big] 
-\frac{\mu_i^2}2   \frac{ ( U^i_x)^2 }{U^i_{xx}}
+\frac12 U^i_{xx} \Big\{ - \big({\theta_i \sigma_i \widehat \sigma}\big)^2\Big\}
\Big\}
\\
= & 
\frac{ U^i_{xx}}2 \frac1{\nu_i^2+\sigma_{i}^2}
\Big| \pi_t^i(\nu_i^2+\sigma_{i}^2)  - \Big(\sigma _{i} \theta_i \widehat \sigma - \mu _{i} \frac{U^i_x}{U^i_{xx}}\Big)  \Big|^2,
\end{align*}
which results in \eqref{eq:SDEafterItoWentzell and choice of Ut-average-n-1}.
\end{proof}

\begin{proof}[of Equation \eqref{eq:generalSDEafterItoWentzell-MFG}]
We take up the drift of \eqref{eq:generalSDEafterItoWentzell-MFG} and we have just by re-organizing the terms

\begin{align*}
0   
&= U_t(Z^\pi_t ,t)+ U_x(Z^\pi_t ,t) \big(\mu \pi_t - \theta \overline{\mu \pi_t} \big)
+\frac12 U_{xx}(Z^\pi_t ,t) \Big( (\nu \pi_t )^2 + \big(\sigma \pi_t - \theta \overline{ \sigma \pi_t } \big)^2\Big)
\\
&
=
\Big( U_t  - U_x  \theta \overline{\mu \pi_t} +\frac 12 U_{xx} \theta^2 (\overline{ \sigma \pi_t })^2 \Big)
\\
&\qquad \qquad 
+\frac12 \frac{U_{xx}}{(\nu^2+\sigma^2)}  \Big( \pi_t^2 (\nu^2+\sigma^2)^2
-2 \pi_t (\nu^2+\sigma^2) \Big\{ \theta \sigma  \overline{ \sigma \pi_t } - \mu  \frac{U_x}{U_{xx}} \Big\}   \Big)
\end{align*}
We recall the optimal strategy given by \eqref{eq:optimstrategy-MFG-vAUXILIARY}, where we complete the square inside the $U_{xx}$ term in the SPDE above we have

\begin{align*}
0   
&=
\Bigg\{    U_t  
       + U_x \cdot\Big( \mu \frac{\theta \sigma  \overline{ \sigma \pi_t }}{(\nu^2+\sigma^2)} -\theta \overline{\mu \pi_t} \Big)+
\frac 12 U_{xx} \cdot \theta^2 (\overline{ \sigma \pi_t })^2 \Big(1- \frac{\sigma^2}{\nu^2+\sigma^2}\Big) 
\\
&
\qquad -\frac12 \frac{\mu^2}{(\nu^2+\sigma^2)}   \frac{(U_x)^2}{U_{xx}}\Bigg\}
+\frac12 \frac{U_{xx}}{(\nu^2+\sigma^2)}  \Big| \pi_t (\nu^2+\sigma^2)
-\Big( \theta \sigma  \overline{ \sigma \pi_t } - \mu  \frac{U_x}{U_{xx}} \Big)   \Big|^2
\end{align*}
Under the CARA condition $U_x/U_{xx}=-\delta$ and the choice of the optimal strategy, the remaining drift must zero-out. We then have
\begin{align*}
U_t &=  -\frac{U_x}{2(\nu^2+\sigma^2)} \cdot\Big( \mu \theta \sigma  \overline{ \sigma \pi_t } +{\delta} {\mu^2} \Big) 
+U_{xx} \frac{(\theta\,\sigma\, \overline{ \sigma \pi_t })^2}{2(\nu^2+\sigma^2)}
-\frac 12 U_{xx} \cdot  (\theta\, \overline{ \sigma \pi_t })^2
+U_x \Big( \theta \overline{\mu \pi_t}\Big). 
\end{align*}
\end{proof}

\normalsize

\end{document}